\documentclass[runningheads]{llncs}
\usepackage[T1]{fontenc}
\usepackage{graphicx}
\PassOptionsToPackage{hyphens}{url}
\usepackage[hidelinks]{hyperref}
\usepackage{color}

\let\anonSubmission\undefined 

\usepackage{algorithmic}
\usepackage{textcomp}
\usepackage{xcolor,mdframed}

\usepackage[nolist, nohyperlinks, printonlyused]{acronym} 
\usepackage[draft]{tikzpeople}
\usepackage[autolanguage,np,boldmath]{numprint}
\usepackage{url}
\usepackage{underscore}

\usepackage{listings}
\usepackage{makecell}
\usepackage{multirow}

\usepackage[inline]{enumitem}

\usepackage{booktabs}

\usepackage{balance}

\usepackage[]{fancyhdr}

\usepackage{tikz}
\usetikzlibrary{positioning, shapes, arrows, fit, trees, matrix, chains, calc, arrows.meta, decorations.pathreplacing,calligraphy}
\tikzstyle{data} = [semithick,rectangle, minimum width=3em, minimum height=1em, text centered, draw=black,align=center]
\tikzstyle{process} = [semithick,rectangle, rounded corners, minimum width=3em, minimum height=1em, text centered, draw=black,align=center]
\tikzstyle{comm} = [thick,<->,>=stealth,sloped]
\tikzstyle{comms} = [thick,->,>=stealth,sloped]
\tikzstyle{commr} = [thick,<-,>=stealth,sloped]
\tikzstyle{energy} = [thick,<->,>=stealth,densely dotted,sloped]
\tikzstyle{contract} = [<->,>=stealth,densely dashed,sloped]
\tikzstyle{arrow} = [thick,->,>=stealth]
\tikzstyle{darrow} = [thick,<->,>=stealth]
\tikzstyle{line} = [thick,-,>=stealth]
\tikzstyle{box} = [draw, dotted]
\tikzstyle{actor} = [semithick,rectangle, minimum width=3em, minimum height=2em, text centered, draw=black,align=center, text width=7em]
\tikzstyle{concept} = [semithick,rectangle, minimum width=6em, minimum height=2em, text centered, draw=black,align=center, text width=11em]
\tikzstyle{backend} = [cloud, cloud puffs=9.7, cloud ignores aspect, semithick, minimum width=2em, minimum height=1em, text centered, draw=black,align=center,fill=white, inner xsep=0.05em, inner ysep=0.05em]

\usepackage[nameinlink,capitalise]{cleveref}      
\crefname{lstlisting}{listing}{listings}
\Crefname{lstlisting}{Listing}{Listings}

\newlist{funcreq}{enumerate}{2}     
\setlist[funcreq]{}
\setlist[funcreq,1]{label=FR\arabic*} 
\setlist[funcreq,2]{label*=.\arabic*} 
\crefname{funcreqi}{Funktional Requirement}{Funktional Requirements} 
\crefname{funcreqii}{Funktional Requirement}{Funktional Requirements} 

\newlist{secreq}{enumerate}{2}     
\setlist[secreq,1]{label=SR\arabic*} 
\setlist[secreq,2]{label*=.\arabic*} 
\crefname{secreqi}{Security Requirement}{Security Requirements} 
\crefname{secreqii}{Security Requirement}{Security Requirements} 

\newlist{privreq}{enumerate}{2}     
\setlist[privreq,1]{label=PR\arabic*} 
\setlist[privreq,2]{label*=.\arabic*} 
\crefname{privreqi}{Privacy Requirement}{Privacy Requirements} 
\crefname{privreqii}{Privacy Requirement}{Privacy Requirements} 

\newlist{subreq}{enumerate}{1}     
\setlist[subreq,1]{label*=\arabic*} 

\newlist{requirement}{enumerate}{1}     
\setlist[requirement,1]{label=R\arabic*} 
\crefname{requirementi}{Requirement}{Requirements} 

\makeatletter
\newcommand{\Git}{\ifdefined\anonSubmission
		\textit{The GitLab URL would not be anonymous and is thus replaced with an OSF link for review:}
		\url{https://osf.io/9a58b/?view_only=b7a9aa9520ae48daaddd67b5fa0a34c4}
		\else
		\url{https://code.fbi.h-da.de/seacop/SSI-PnC-Tamarin}
		\fi}
\makeatother

\begin{document}

\title{Self-Sovereign Identity for Electric Vehicle Charging}

\ifdefined\anonSubmission
\author{}
\institute{}
\else
\author{Adrian Kailus\inst{1} \and
Dustin Kern\inst{2} \and
Christoph Krauß\inst{2}
}
\institute{DB Systel GmbH, Germany\\
	\email{a@kailus.dev} \and
	Darmstadt University of Applied Sciences, Germany\\
	\email{\{dustin.kern,christoph.krauss\}@h-da.de}}
\fi

%
%


\maketitle

\begin{abstract}
\acp{EV} are more and more charged at public \acp{CP} using \ac{PnC} protocols such as the ISO~15118 standard which eliminates user 
interaction for authentication and authorization. Currently, this requires a rather complex \ac{PKI} and enables driver 
tracking via the included unique identifiers. In this paper, we propose an approach for using \acp{SSI} as trusted credentials 
for \ac{EV} charging authentication and authorization which overcomes the privacy problems and the issues of a complex 
centralized \ac{PKI}. Our implementation shows the feasibility of our approach with ISO~15118. The security and privacy of the proposed approach is 
shown in a formal analysis using the Tamarin prover.

\end{abstract}

\begin{acronym}[DHCPv6]
	\acro{ACN}{Adaptive Charging Network}
	\acro{AIS}{Artificial Immune System}
	\acro{AMI}{Advanced Metering Infrastructure}
	\acro{AUC}{Area Under the Curve}
	\acro{BMS}{Battery Management System}
	\acro{CA}{Certificate Authority}
	\acrodefplural{CA}[CAs]{Certificate Authorities}
	\acro{CAN}{Controller Area Network}
	\acro{CCH}{Contract Clearing House}
	\acro{CL}{Camenisch-Lysyanskaya}
	\acro{CP}{Charge Point}
	\acro{CPO}{Charge Point Operator}
	\acro{CPS}{Certificate Provisioning Service}
	\acro{DAA}{Direct Anonymous Attestation}
    \acro{DID}{Distributed Identifier}
	\acro{DNS}{Domain Name System}
	\acro{DoS}{Denial of Service}
	\acro{EMAID}[eMAID]{e-Mobility Account Identifier}
	\acro{EMSP}[eMSP]{e-Mobility Service Provider}
	\acro{EV}{Electric Vehicle}
	\acro{FDI}{False Data Injection}
	\acro{FPR}{False Positive Rate}
	\acro{HAN}{Home Area Network}
	\acro{HEMS}{Home Energy Management System}
	\acro{HSM}{Hardware Security Module}
	\acro{IDS}{Intrusion Detection System}
	\acro{IDPS}{Intrusion Detection and Prevention System}
	\acro{IoT}{Internet of Things}
	\acro{Mad}{Manipulation of demand}
	\acro{MadIoT}{Manipulation of demand via IoT}
	\acro{MitM}{Man-in-the-Middle}
	\acro{MV}{Medium Voltage}
	\acro{NAN}{Neighborhood Area Network}
	\acro{OEM}{Original Equipment Manufacturer}
	\acro{OCPI}{Open Charge Point Interface}
	\acro{OCPP}{Open Charge Point Protocol}
	\acro{OSCP}{Open Smart Charging Protocol}
	\acro{PCID}{Provisioning Certificate Identifier}
	\acro{PKI}{Public Key Infrastructure}
	\acro{PLC}{Power Line Communication}
	\acro{PnC}{Plug-and-Charge}
	\acro{R2L}{Remote to Local}
	\acro{RFID}{Radio Frequency Identification}
	\acro{ROC}{Receiver Operating Characteristic}
	\acro{RTT}{Round Trip Time}
	\acro{SCADA}{Supervisory Control and Data Acquisition}
	\acro{SMGW}{Smart Meter Gateway}
	\acro{SoC}{State of Charge}
	\acro{SSI}{Self-Sovereign Identity}
	\acrodefplural{SSI}[SSIs]{Self-Sovereign Identities}
	\acro{SVM}{Support Vector Machine}
	\acro{TLS}{Transport Layer Security}
	\acro{TPM}{Trusted Platform Module}
	\acro{TPR}{True Positive Rate}
	\acro{U2R}{User to Root}
	\acro{V2G}{Vehicle to Grid}
	\acro{WAN}{Wide Area Network}
\end{acronym}

\acresetall

\section{Introduction}

\ac{PnC}, e.g., using the standard ISO~15118, enables \acp{EV} to charge without user interaction at public \acp{CP} 
operated by a \ac{CPO}. The \ac{EV} stores relevant data such as contract credentials and automatically performs all 
necessary steps to start a charging session, e.g., authentication, authorization, and negotiation of charging parameters. No
\acsu{RFID} cards or smartphone apps are required anymore. To enable this, ISO~15118 defines a complex \ac{PKI} and 
uses a unique identifier to identify the user or actually the user's personal charging contract. The charging contract
is the basis for billing of \ac{PnC} sessions and is concluded between an \ac{EV} user and an \ac{EMSP}. 

The complex \ac{PKI} architecture of ISO~15118 requires all entities to operate central (sub-) \acp{CA}. These 
entities include \acp{CPO} and \acp{EMSP} but also \acp{OEM} and a \ac{CCH}.
\acp{OEM} produce \acp{EV} and the \ac{CCH} enables roaming 
services for charging at \acp{CP} from different operators. Furthermore, the Root \acp{CA} are possible single points of 
failure. The unique identifier of the charging contract, called \ac{EMAID}, enables user tracking which raises  
privacy issues. By analyzing movement profiles, user habits or even the health status may be deduced, e.g., if the vehicle is 
regularly charged at a hospital.

To overcome the issues of centralized systems such as \acp{PKI} or identity pro\-vi\-ders, \acp{SSI} gained a lot of attention 
in the last years. \ac{SSI} provides a digital identity and enables users to control the information they disclose to prove their identity and to 
protect their privacy.

In this paper, we propose an approach for using \acp{SSI} as trusted credentials for \ac{EV} charging authentication and 
authorization. Our approach solves the issues of complex centralized \ac{PKI} and protects against linking multiple authentication processes.
The contributions of this paper are as follows:
\begin{enumerate*}[label=(\roman*)]
	\item Concept for the secure integration of \ac{SSI} into ISO~15118 with privacy-preserving charging authentication/authorization.
	\item Proof-of-concept implementation showing minor additional overhead and easy integration into existing systems.
	\item Formal security and privacy analysis in the symbolic model using the Tamarin prover~\cite{meier2013tamarin}.
	\item Publishing the used Tamarin models (cf. \cref{sec:evaluation:t}) for reproducibility of the automated proofs and reusability of used modeling concepts in related work.
\end{enumerate*}

The remainder of the paper is structured as follows: 
\cref{sec:background} describes necessary background to understand our approach. Related work is discussed in \cref{sec:rel}. 
In \cref{sec:req}, we present identified requirements for 
our concept which is introduced in \cref{sec:concept}. Our prototypical implementation is described in 
\cref{sec:implementation}, followed by the security, privacy, and practical evaluations in \cref{sec:evaluation}. Finally, we conclude 
the paper and discuss future work in \cref{sec:conclusion}.

\section{Background}\label{sec:background}

In this section, we describe background on e-mobility and \ac{SSI}. The focus is on the certificate-based authentication 
which we replace with \ac{SSI} credentials. 

\subsection{E-Mobility} \label{sec:emob}

\begin{figure}[]
	\centering
	\includegraphics[width=0.56\linewidth]{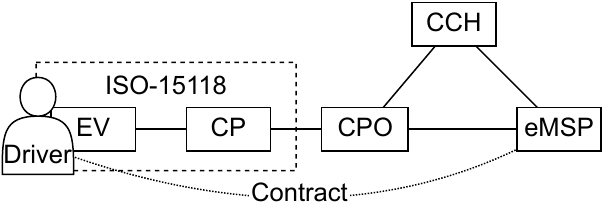}
	\caption[Architecture Overview]{Architecture Overview (cf. \cite{ELAADNL2017})}
	\label{fig:background:evachitectureoverview}
\end{figure}

\cref{fig:background:evachitectureoverview} shows a simplified e-mobility architecture according to the ISO~15118 standard. 
There exist two editions 
of the standard, the first edition ISO~15118~\cite{iso1,iso2} and the second edition ISO~15118-20~\cite{iso2-ed2} which 
brings some security improvements. Our solution can be applied to both versions.

An \ac{OEM} manufactures the \ac{EV} (not shown), provides some initial credentials to the \ac{EV}, and sells it to the new owner. The 
owner concludes a contract with an \ac{EMSP} for charging at public \acp{CP} which are operated by a \ac{CPO}. 
The initial credential from the \ac{OEM} are used by the \ac{EV} to request contract credentials from the \ac{EMSP}.
A \ac{CPS} establishes trust in the contract credentials provided by the \ac{EMSP}.
The \ac{EV} stores and uses the contract credentials for \ac{PnC} authorization and billing during a charging session with a \ac{CP}. 
The communication between \ac{EV} and \ac{CP} is secured with \acsu{TLS}. The first edition of ISO~15118 uses 
unilateral \ac{TLS} authentication of the \ac{CP} and challenge-response-based authentication of the \ac{EV} inside the \ac{TLS} channel. The second edition uses mutual authentication with a vehicle certificate installed by the \ac{OEM} in addition to 
the challenge-response-based \ac{EV} authentication.

ISO~15118 requires multiple certificates and defines a rather complex \ac{PKI}. The \ac{PKI} consists of four\footnote{We 
omit the part for private environments since it is not relevant for our work.} parts for \ac{CPO}, \ac{OEM}, \ac{EMSP}, and 
\ac{CPS}. All \acp{PKI} consist of up to two sub-CAs below a root CA. The root CA for \ac{CPO}- and \ac{CPS}-\ac{PKI} is the 
V2G root \ac{CA} which usually also certifies the sub-CAs of \ac{EMSP} and \ac{OEM} via cross-signing. The 
\ac{EMSP}-\ac{PKI} and \ac{OEM}-\ac{PKI} are always certified by their own root CAs. 

The \ac{CPO}-\ac{PKI} is used for issuing certificates for \acp{CP} which are used for \ac{CP} authentication 
in the \ac{TLS} handshake. 

The \ac{OEM}-\ac{PKI} is used to issue the \ac{OEM} provisioning certificate which includes the unique identifier 
\ac{PCID}. The \ac{OEM} provisioning certificate is used as initial trust anchor for installing the contract credentials. In 
case the second edition ISO~15118-20 is used, additionally, a vehicle certificate is issued for \ac{EV} authentication in 
the \ac{TLS} handshake.

The \ac{EMSP}-\ac{PKI} is used to generate the contract certificate after concluding a contract with an \ac{EV} owner. The 
\ac{EMSP} generates contract certificate data which consists of a private key and the contract certificate (including the 
corresponding public key, a unique identifier called \ac{EMAID}, and additional information). This data is installed when 
the \ac{EV} is first connected to a public \ac{CP}. The private key is encrypted with the public key of the 
\ac{OEM} provisioning certificate to ensure that only the specific \ac{EV} can access this key. 

Finally, the \ac{CPS}-\ac{PKI} is used for generating certificates which are used by a \ac{CPS} to sign contract 
certificate data generated by the \ac{EMSP}. An \ac{EV} can verify the signature and the certificate chain up to the known 
V2G root \ac{CA}. Thus, the verifier does not need to know the \ac{EMSP} root \ac{CA}.

In addition to the complexity of the PKI, there is another issue in ISO~15118 namely the lack of privacy protection. 
Currently, a lot of information, arguably not required for operation, is disclosed to entities such as \acp{CPO}, \acp{CCH}, 
and \acp{EMSP} \cite{kern2022integrating}. For example, it would not be necessary to send the exact time and \ac{CP} location of a charging session to 
the \ac{EMSP} or the \ac{EMAID} to the \ac{CPO}.

\subsection{\acf{SSI}} \label{sec:ssi}
A \acf{SSI} allows a user to create and fully control a digital identity without requiring centralized infrastructures or 
identity providers. The user can also control how personal data is shared and used by another party via a 
decentralized path. After an information is verified by an issuer (e.g., a university verifying a degree), a verifier (e.g., a 
company) can always trust that information to be true. Subsequently, the information holder (e.g., a student) does not need to 
provide the full information to the verifier to prove its identity. This is achieved using verifiable credentials (standardized 
by the W3C \cite{w3cVCdatamodel}), the distributed identity protocol, and a distributed ledger technology (which is mostly a 
blockchain). The information holder registers an information identifier at a ledger, which is verified by an issuer, and the 
verifier can trust this information. In the following, we introduce the most relevant terms for our work.

\paragraph{Verifiable Claims} 
In \ac{SSI}, the essence is that a counterpart can rely on a claim without having control over the content of the claim. 
Here, a distinction must be made between a Claim and a Verifiable Claim. First, a claim is simply a statement about a fact 
that anyone could make and without being verifiable. For example, it could be stated that Alice is a graduate of a certain 
university. 
However, for this statement to become a 
Verifiable Claim, the signature of an issuer may be added to it.
Alternatively, zero-knowledge cryptography may be used in a privacy-preserving manner to indirectly prove that a claim is covered by a valid verifiable credential \cite{w3cVCdatamodel}.

\paragraph{Verifiable credentials}
A collection of claims together with an identifier and metadata such as the issuer, expiration date, terms of use, and keys form 
a credential. Credentials are comparable to conventional ID documents, which likewise bundle a number of statements. 
Multiple credentials can be combined into one profile.\footnote{Combining credentials, 2018, 
\url{github.com/w3c/vc-data-model/issues/112}}

\paragraph{Decentralized Identifiers}
Identifiers that can be resolved to a \ac{DID} Document\footnote{DID resolution, W3C, 2021, 
\url{w3c-ccg.github.io/did-resolution/}} and do not require a central registration authority to be created. The \ac{DID} 
Document, which can only be modified by the \ac{DID} Controller, can contain information about public keys, verification methods, 
the controller, and authentication methods, among other things. The \ac{DID} Controller also defines the subject of the \ac{DID}, 
e.g., a person or organization. Specific sections in a \ac{DID} document can be referenced by the respective \ac{DID} URL. Both the 
\ac{DID} and the \ac{DID} Document are stored in a \textit{Verifiable Data Registry} (e.g., a distributed ledger) and their 
combination is called a \ac{DID} Record. The public keys of a \ac{DID} enable encrypted communication with the owner of the 
\ac{DID}. To do this, a communication partner can either use a DID Record they got from the other party or look up the public 
keys in the \textit{Verifiable Data Registry} \cite{Reed2021DecentralizedDraft}.

\paragraph{DID Auth}
There are 10 different architectures to authenticate an identity holder using different transports for the challenge-response 
cycle \cite{Sabadello2018IntroductionAuth}. The main focus is to let an identity holder prove to have control over a \ac{DID}. 
Authentication can be unilateral or bilateral, with both parties demonstrating control over their own \ac{DID}. This may also 
involve the exchange of Verifiable Credentials if required by the use case. There are three ways to combine \ac{DID} Auth with 
Verifiable Credentials: \ac{DID} Auth and the Verifiable Credentials are exchanged separately (in that 
order); The Verifiable Credentials are part of \ac{DID} Auth and represent an optional field in the authentication protocol 
or finally, \ac{DID} Auth can be considered a special case of a Verifiable Credential, with a claim "I am me". The authentication 
process is based on a challenge-response cycle where the relying party authenticates the identity holder using, for example, a 
cryptographic signature.

\section{Related Work}\label{sec:rel}

The increasing integration of information and communication technology into vehicles enables automated tracking of vehicles which threatens the 
privacy of drivers and passengers \cite{bradbury2020privacy}. \cite{langer2013privacy} discusses privacy issues for electric mobility and 
\cite{han2016privacy} privacy challenges for \ac{EV} charging.

Several approaches for security and/or privacy in \ac{EV} charging have been proposed. In \cite{li2014portunes,li2016portunes+}, an \ac{EV} authentication protocol for 
contactless charging (i.e., using charging pads integrated into the road) using pseudonyms is proposed. An architecture for privacy-preserving 
contract-based charging and billing of \acp{EV} using ISO~15118 is presented in \cite{hofer2013popcorn}. A formal analysis and improvements of this 
architecture are presented in \cite{fazouane2015formal}. A privacy-preserving solution for roaming \ac{EV} charging and billing based on smart cards 
is proposed in \cite{mustafa2014roaming}. The solutions presented in \cite{kern2022integrating,zelle2018anonymous,zhao2015secure} all require a 
\ac{TPM} to realize a \ac{DAA} scheme for \ac{EV} authentication. Using a \ac{TPM} for protecting credentials but without privacy protection is 
proposed in \cite{fuchs2020hip,fuchs2020b,fuchs2020a,fuchs2020hip20}. All approaches still require a complex \ac{PKI}.

Some work exists that seeks to address the issues around privacy and user profiling when charging \acp{EV} via the implementation of a new, anonymous 
payment channel. This often involves a blockchain solution that promises anonymous payment processing and a decentralized infrastructure. 
The authors of \cite{Firoozjaei2019EVChain:Charging}, for example, present a solution in \cite{Firoozjaei2019EVChain:Charging} where payment for charging electricity is handled through multiple blockchains. A main block\-chain negotiates transactions between the operator and the \acp{CP}, and on sub-block\-chains, multiple customers join together to form credit sharing groups in which individual payments cannot be linked to the buyer of the credits. Here, the degree of anonymity is measured using K-anonymity, which quantifies the group size from which a user is indistinguishable. The main blockchain is connected to the sub-blockchains via a \textit{bridge} role that communicates with credit buyers.
The authors of \cite{Xu2021EVchain:Vehicles} also present a blockchain-based solution for charging \acp{EV} in \cite{Xu2021EVchain:Vehicles}, which is also 
based on \textit{K-Anonymity}. Their approach uses a distributed \ac{PKI} that separates user registration and verification across two blockchains. 
Payment here is handled via smart contracts. 
In \cite{knirsch2018privacy}, a blockchain-based approach for privacy-pre\-serv\-ing selection of a \ac{CP} based on tariff options and travel 
distance is presented. 
The authors of \cite{JonghoWon2018DecentralizedInternet-of-Things} propose the implementation of a blockchain-based \ac{PKI} for \ac{IoT} 
\cite{JonghoWon2018DecentralizedInternet-of-Things} and demonstrate the feasibility and efficiency of such an IoT PKI through a prototype 
implementation and experiments. The PKI network is based on Emercoin 15 and uses a proof-of-stake consensus algorithm.

Some work already considers the use of \ac{SSI} for \ac{EV} charging.
The authors of \cite{richter2021exploring} provide a high-level analysis of the potential benefits that an \ac{SSI} solution can  bring to \ac{EV} charging.
However, no detailed concept is proposed and details on, e.g., the integration into existing \ac{EV} charging processes or the resulting overhead are not analyzed.
Similar to our work is the approach of \cite{Hoess2022WithOW}, which also uses \ac{SSI} for decentralized eRoaming. 
However, this concept differs from our ISO~15118 extension and makes use of the user's smartphone instead of allowing for \ac{PnC}-based \ac{EV} authentication without user interaction. 
Also, no implementation is developed and a detailed analysis of performance overhead and security is provided.

In contrast to related work, our work presents a novel solution for the integration of \ac{SSI} into the \ac{EV} charging ecosystem.
We consider the integration into existing protocols and process to enhance the potential usability of the solution as much as possible.
Additionally, we provide a performance analysis based on a proof-of-concept implementation as well as a formal security and privacy analysis using the Tamarin prover \cite{meier2013tamarin}.

\section{System Model and Requirement Analysis}\label{sec:req}
The following section outlines the scope of this work, defines an attacker model, and lists and discusses the concept requirements,  which are grouped into three categories: Functional Requirements (FR), Security Requirements (SR), and Privacy Requirements (PR).

\subsection{Scope}
\label{sec:req:scope}
Among other things, the \ac{PnC} process maps a bidirectional authentication between \ac{CP} and \ac{EV} to trust the existence of a contractual relationship and to rule out malicious actors. These authentications in ISO 15118 are based on a common \ac{PKI}, which is used, among other tasks, to authorize a vehicle for a charging process, to authenticate the charging infrastructure, or to establish the TLS connection. 
In an all-encompassing extension of traditional authentication via \ac{PKI} and its certificates, both authentications would therefore be replaced, including their use for the \ac{TLS} connection and the \textit{metering} messages during the charging loop, while this work is limited to the authentication process of the contract information provided by the \ac{EV} to the \ac{CP}. 


\subsection{Attacker Model}
\label{sec:req:attackermodel}
In order to make the concept viable against possible attacks and vulnerabilities, an attacker model is set up in the following.

Classic attacker models, such as the Dolev-Yao Model \cite{dolev1983security}, 
outline malicious network participants capable of intercepting network communications, sending and modifying messages. However, we assume that basic cryptographic primitives and implementations hold \cite{monteuuis2018attacker}. 


Additionally, we consider threats to the system's privacy. The centralized approach to certificate validation makes users traceable and their personal data vulnerable to attack by any of the actors. This threat is increased in case one of the actors is compromised by an attacker or stops following the agreed protocol to obtain additional information. While such \textit{malicious operators} pose a major threat, the danger posed by such \textit{malicious operators} is limited \cite{paverd2014modelling}.
This is mainly due to the fact that operators have to comply with legal regulations and maintain their image to the public. 
Taking this into account, the \textit{Honest-but-Curious Operator} is described below (cf. \cite{kern2022integrating}).

Above all, the \textit{Honest-but-Curious Operator} does not want to create a malicious impression to the outside world by deviating from the agreed protocols. Since involved in the process, such operators use all information available to them to ultimately derive additional benefit from it. In the \ac{PnC} context, potentially \textit{Honest-but-Curious Operators} can include the \acp{CPO}, \acp{EMSP} and the \ac{CCH}.
At this point it is assumed that several operators do not accumulate their available information to draw a more comprehensive data picture, since this is opposed to the competition relationship among operators and should additionally be prevented by regulations. Ultimately, the regulation of operators is beyond the control of this concept.


\subsection{Functional Requirements}
\label{sec:req:functional}

In order to ensure user-friendliness and to allow for an easy integration of the solution into existing protocols and processes, we define several functional requirements.
The requirements ensure that features of the original ISO~15118 can be supported by the new concept.
For example, in order for the vehicle to authenticate itself at the charging stations with its contract information, a process must be defined for contract installations which provide the vehicle with the necessary information.
In order to uniquely associate a driver's contract with the vehicle, the vehicle must be uniquely identifiable during the installation process. 
In order to ensure that the solution is user-friendly, any additional overhead should remain acceptable.
Functional Requirements (FR) are listed in the following:

\begin{funcreq}[resume]
 \item Vehicle charging as well as contract installation should still be possible without further user interaction, since this is the concept of \ac{PnC}. \label{requirement:interaction}
 \item Contract authentication via \ac{SSI} should be negotiable as an option to the existing authentication methods. \label{req:optional} 
 \item All \ac{SSI} roles should be able to be taken by an actor from the ISO~15118 ecosystem. \label{req:roles01} In \ac{SSI}, the credential verification process principally covers three roles: the Issuer, the Holder and the Verifier, which must be uniquely applied to an entity in the \ac{PnC} context for each authentication.
 \item The vehicle should continue to manage the necessary authentication information itself (in a wallet). \label{req:wallet}



 \item All Contract Issues from all Issuers should fit an agreed schema baseline. \label{req:iss:schema}
 \item As in ISO-15118, it should be possible to delay the installation of the contract information until the first charging process. \label{req:installtime}

 \item The charging station should relay communication from the vehicle to the other actors in case the vehicle cannot use cellular. \label{req:comm:tunnel}

	\item The additional computational- and communication overhead of a \ac{SSI}-based solution should be minor. \label{req:perf}
\end{funcreq}

\subsection{Security and Privacy Requirements}\label{sec:req:sec}
The non-functional requirements for the concept are listed and explained below. This includes Security Requirements (SR) and Privacy Requirements (PR).
The security requirements focus on providing secure authentication for the actors involved in relevant processes (setup, credential installation, charging, billing):

\begin{secreq}[resume]
    \item The setup proceeds of the solution should be secure (e.g., the setup of \acp{EV} with provisioning credentials or the setup of \acp{EMSP} as issuers of verifiable credentials). That is, all relevant parties should be securely authenticated to enable trust between the parties. \label{req:iss:trust}

 \item During the contract credential installation the \ac{EMSP} should be able to trust in the originality of the vehicle, similarly to the \ac{OEM} provisioning certificate in ISO-15118, which is installed during vehicle production. That is, the \ac{EV} should securely authenticate itself towards the \ac{EMSP} during the credential installation process. \label{req:issue:trust}



 \item The \ac{CP}/\ac{CPO} should be able to trust the \ac{EV}'s provided contract information. That is, the \ac{EV} should securely authenticate itself towards a \ac{CP} before the start of a charging process. \label{req:cont:trust}
 \item The contract information should allow the \ac{EMSP} to associate an invoice from a \ac{CPO} with a contract. That is, the \ac{EV}'s charge authentication data should securely authenticate the \ac{EV}'s contract towards the \ac{EMSP} for billing. \label{req:cont:payment2}
\end{secreq}
The privacy requirements focus non-traceability and non-linkability of \ac{EV} users:

\begin{privreq}[resume]
 \item During the authentication process no information should be exchanged that makes the user traceable to either a \ac{CPO}, \ac{CCH} or an \ac{EMSP}, preventing the creation of a user's movement profile (\textit{non-traceability}). \label{req:char:traceability}
 \item A specific \ac{CPO}, \ac{CCH} or \ac{EMSP} should not be able to associate multiple charging operations with individual users (\textit{non-linkability}). \label{req:char:linkability}
\end{privreq}
Notably, traceability and linkability of \ac{EV} users by their \ac{EMSP} is feasible due to payment processing via traditional payment methods.
This problem may be solved by using smart contracts (cf. \cite{Xu2021EVchain:Vehicles}), which is out-of-scope for this paper.

\section{\acs*{SSI} Concept}\label{sec:concept}
In the following, a concept for integrating an \ac{SSI}-based solution into the ISO~15118 authentication process is developed, including an architectural overview and the message sequences of the communication between the actors.
The main challenge is in the specific combination of the different \ac{SSI} concepts (cf. \cref{sec:ssi}) such that actors, processes, and features of the existing \ac{EV} charging architecture can still be supported while also designing the concept in a way that enables the (Tamarin-based) symbolic verification of the strong security and privacy requirements (cf. \cref{sec:req:sec}).

\subsection{Concept Overview}
\label{sec:concept:conceptoverview}
In this specific scenario, the already existing parties of ISO~15118 are sufficient to map all three roles \textit{Holder}, \textit{Verifier}, and \textit{Issuer} of the \ac{SSI} process. 

The Holder and the Verifier of the contract authentication process are easy to identify in the \ac{PnC} context: The Holder is the actor in possession of the contract information. This data could be stored either in a wallet on the driver's smartphone, along with other credentials, or in the \ac{EV} in the form of an on-board wallet. The first option would require driver consent each time information is accessed from the wallet, similar to 
\cite{Lux2020Distributed-Ledger-basedCredentials}. Since the main goal of \ac{PnC} is to enable vehicle charging without further user interaction, it is preferable to install the wallet in the \ac{EV}. This also eliminates the need to communicate with the driver's smartphone.
Since the verifier needs to authenticate the contracts, this role is taken by the \ac{CP}, which is already performing this task in ISO~15118. 

The issuer first needs access to the original contracts to authenticate them as credentials. This condition applies only to the \ac{EMSP}, with each \ac{EMSP} having access solely to the contracts of its clients. Furthermore, the verifiers, i.e., the \acp{CP}, should be able to trust the issuer. Since the \acp{CP} already had to trust the \acp{EMSP} in the conventional ISO~15118, this condition is also met.

To grant multiple issuers write permissions on the Ledger to create documents like \textit{Credential Definitions} or \textit{Credentials}, an additional instance is needed that can give these permissions to the different issuers - the \textit{Steward}.

\begin{figure}[htbp]
 \centering
 \includegraphics[width=0.8\linewidth]{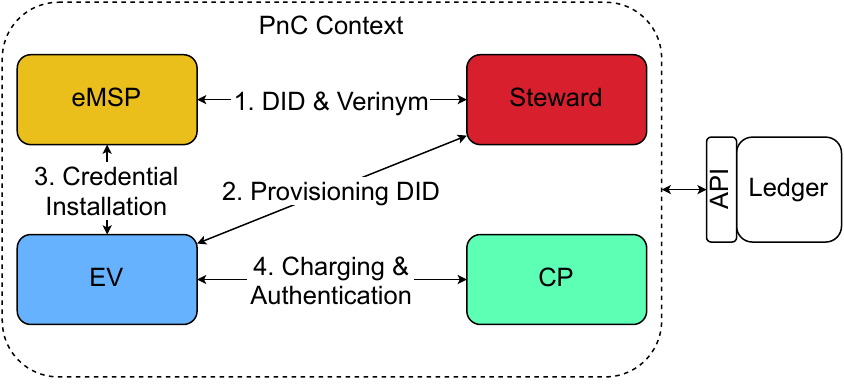}
 \caption{Architecture Overview}
 \label{fig:concept:achitectureoverview}
\end{figure}

\cref{fig:concept:achitectureoverview} shows how these four actors interact for charging authentication in the overall system. 
Initially, only the steward is authorized to write to the ledger which reduces the number of first-level write permissions. The steward grants second-level write permissions to new \acp{EMSP} later on.
The steward writes these permissions to the ledger in the form of a verinym (step 1), which enables the \ac{EMSP} to authenticate its contracts.
A verinym is associated with the legal identity of the identity holder 
\cite{HyperledgerIndy}.
Thus, the legal entity of the \ac{EMSP} that enters into the contracts with the customers is associated with the identity on the ledger that has write permissions for the credentials of those same contracts. 

In step 2, a \textit{Provisioning \ac{DID}} is created for the vehicle. This is done before the vehicle is sold. This \textit{Provisioning \ac{DID}} is necessary to be able to link a specific vehicle to a contract later on. Furthermore, with the help of the public key of a \ac{DID}, it is always possible for other actors to communicate with its owner in an encrypted way, which will also be helpful later on. Of course, this also applies to all other \ac{DID}s used in the \ac{PnC} context.

Then, in order for the necessary contract information to be authenticated during a charging process, the information must be transferred to the vehicle. This third step can happen once a contract is established and the vehicle has connected to the internet (directly or via a \ac{CP}). Since the vehicle may have wireless, but this is optional, this step can take place sometime after the \textit{Provisioning \ac{DID}} has been created between the conclusion of the contract and the charging process. For this, the vehicle requests the credentials from the respective \ac{EMSP}, which authenticates them on the ledger. 

The vehicle can then authenticate itself to the \ac{CP} during the charging process in the final step 4.
Authentication uses \emph{Anoncreds},\footnote{\url{github.com/hyperledger/indy-sdk/tree/main/docs/design/002-anoncreds}} i.e.,  zero-knowledge proofs with \ac{CL}-based credentials and paring-based revocation \cite{khovratovich2018anonymous}.
In short, the \ac{EV} proofs to the \acp{CP} that it possesses valid contract credentials and that these credentials have not been revoked by the issuer (without revealing the actual credentials).

The following sections describe the changes made to the message sequence of ISO~15118 in order to create a working infrastructure for the transition to \ac{SSI} authentication.


\subsection{Provisioning DID Creation}
\label{sec:concept:contractissue:provdidcreation}


Prior to any charging process, the issuer, in this case the \ac{EMSP}, must be authorized to issue credentials.
That is, the \ac{EMSP} needs write permission to the ledger, which requires publishing its \ac{DID} (containing a public key) to the ledger. Such a \ac{DID} is often called a \textit{Verinym}. The \ac{EMSP} makes a request to the steward, which is authorized to write to the ledger. 
This process is secured based on pre-negotiated secret or public keys. Since both communication partners are legal entities, it can be assumed that there is an agreement between the two in which a secret or public key can be exchanged. 

Anther setup process is the creation of a \textit{Provisioning \acp{DID}} (cf. \cref{fig:concept:provdidcreation}), which is a  prerequisite for linking the contract and the vehicle. This process is described in the following paragraphs:

\begin{figure}[htbp]
	\centering
	\includegraphics[width=0.8\linewidth]{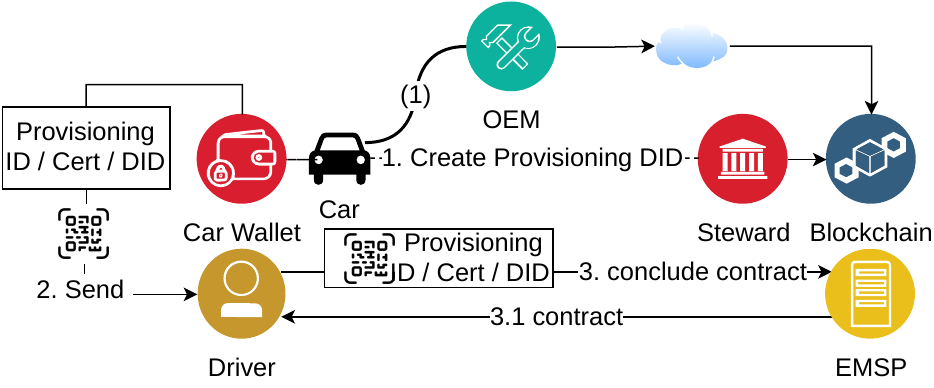}
	\caption{Provisioning \textit{DID} Creation}
	\label{fig:concept:provdidcreation}
\end{figure}

\paragraph{Step 1} The \ac{EV} provisioning process, starts with the production of the vehicle. During this process, the \ac{EV} creates a \textit{Provisioning \ac{DID}}, which enables encrypted communication using the vehicle's public key. A part of the \ac{DID} is the \textit{DID record}, which contains the public information for a given \ac{DID} and must be written to the ledger. 
In short, after connecting to the steward either via the \ac{OEM} or cellular, the \ac{EV} (or \ac{OEM}) starts with sending an \textit{InitNymReq} with a nonce, 
answered by the steward with an \textit{InitNymRes}, containing a \ac{DID} for a key of the steward, the \ac{OEM}'s nonce, a fresh nonce from the steward and the \ac{OEM}'s ID. 
The \textit{InitNymRes} is signed by the steward (with the key corresponding to the \ac{DID}) and encrypted with a public key of the \ac{OEM}. 
The steward's \ac{DID} allows the \ac{EV} to encrypt future messages to the steward, and the nonces are used to ensure replay-protection and subsequently a proof of possession for the \ac{EV}'s \textit{Provisioning \ac{DID}}. 
The \ac{EV} (or \ac{OEM}) creates a \textit{Provisioning \ac{DID}}, decrypts the steward message, verifies the signature, and signs the steward's nonce with the private key of the \textit{Provisioning \ac{DID}}.
The \textit{Provisioning \ac{DID}} (including the corresponding public key) and the signature are sent back to the steward, encrypted with the public key from the steward's \ac{DID}.

\paragraph{Steps 2 and 3}
When the vehicle is purchased, the \textit{Provisioning \ac{DID}}, is passed to the user so that the user can pass the \textit{Provisioning \ac{DID}} to the \ac{EMSP} and negotiate a contract. The handover at the time of concluding a contract with the \ac{EMSP} could be via a QR code sent to the user, who then activates the contract by passing on the \ac{DID}, but other ways are not excluded. Since a potential co-reader does not have the private keys of the \ac{DID}, he cannot prove their possession and cannot succeed in a challenge. This completes the process until the first charging session.

\subsection{Contract Credential Installation}
\label{sec:concept:contractissue:installationprocess}

\begin{figure}[htbp]
 \centering
 \includegraphics[width=0.8\linewidth]{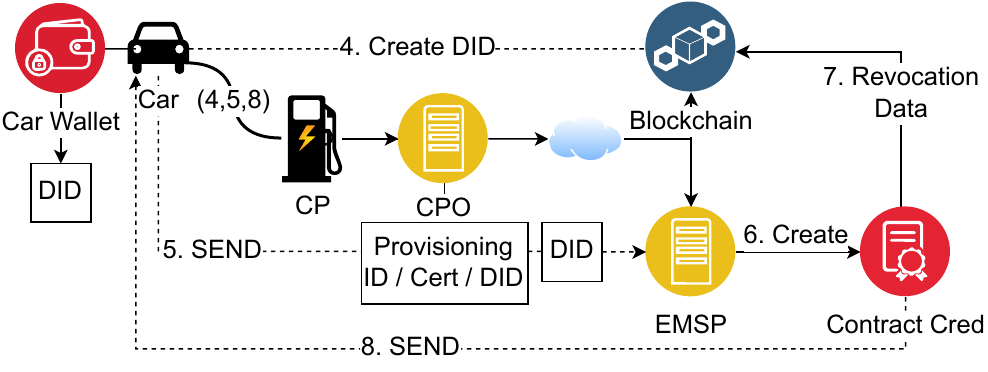}
 \caption{Contract Credential Installation}
 \label{fig:concept:installationprocess}
\end{figure}

The following is an explanation of the general process steps for installing the Contract Credential (cf. \cref{fig:concept:installationprocess}), which requires a \textit{Provisioning \ac{DID}} and an existing contract with an \ac{EMSP}. 
This process is modeled on the Issue Credential Protocol from \cite{HyperledgerAries0036}.

\paragraph{Step 4 and 5}
Contract credentials are required in the vehicle during a charging process. To do this, they must first be created.
As the vehicle may not be able to connect to the Internet, and thus to the ledger and other services, until it is plugged into a \ac{CP} for the first time. 
Once the connection is established, the \ac{EV} starts by sending its \textit{Provisioning \ac{DID}} to the \ac{EMSP}.
The \ac{EMSP} responds with its \ac{DID} and a \textit{Credential Offer}, which includes a nonce and a \textit{Credential Definition ID}. The latter identifies a credential schema, which specifies the structure of all issued credentials (of a certain contact type) by this \ac{EMSP} with all necessary and optional fields, with public keys and a \textit{Revocation Registry}.
The \ac{EMSP}'s response is encrypted for the \ac{EV} based on the \textit{Provisioning \ac{DID}}.

\paragraph{Steps 6, 7, and 8}
If the \ac{EV} agrees to this \textit{Credential Offer}, it 
generates a master secret for the credential. 
The \ac{EV} then creates a blinded master secret for the \textit{Credential Offer} and a correctness proof.
Afterwards, the \ac{EV} builds a \textit{Credential Request} with the blinded master secret and correctness proof and encrypts this request based on the \ac{EMSP}'s \ac{DID}.

The \ac{EMSP} decrypts this \textit{Credential Request} and uses it to create the \textit{Contract Credentials} that an \ac{EV} needs in order to authenticate itself at \acp{CP}. 
Additionally, the \ac{EMSP} updates the revocation information, i.e., the public tails files and the accumulator\footnote{\url{hyperledger-indy.readthedocs.io/projects/hipe/en/latest/text/0011-cred-revocation/README.html}} on the ledger to include the new credential.
This step can optionally include the revocation of old credentials in case a contract has been terminated or the terms of the contract have changed.

The \textit{Contract Credentials} need to be authenticated by an authorized issuer, which can be the \ac{EMSP}, and  contain all billing-relevant information as attributes.
This billing-relevant information, is at least, the \ac{EMSP}'s ID, which is needed by \acp{CP}/\acp{CPO} to identify the \ac{EV} user's \ac{EMSP} for billing purposes.
Additionally, the credential attributes can include any tariff information that may be useful to \acp{CP}/\acp{CPO} (e.g., pricing thresholds or if \ac{V2G} power transfer is supported).
The \Ac{EV} user can always decide which attributes from a \textit{Contract Credential} they want to reveal during a zero-knowledge proof.

The \ac{EV} receives the signed \textit{Contract Credentials} along with the credential revocation information from the \ac{EMSP} encrypted with the public key of the \textit{Provisioning \ac{DID}} via the existing connection in a \textit{CreateContractCredentialRes}.
The \ac{EMSP}'s response additionally includes a symmetric contract key, which is later used to securely authenticate the \ac{EV}'s contract towards the \ac{EMSP} for billing.
The \ac{EV} decrypts and verifies the received data and stores at authentication during charge sessions. 

\subsection{Charging Process and Credential Validation}
\label{sec:concept:contractissue:credentialvalidation}

\begin{figure}[htbp]
 \centering
 \includegraphics[width=0.8\linewidth]{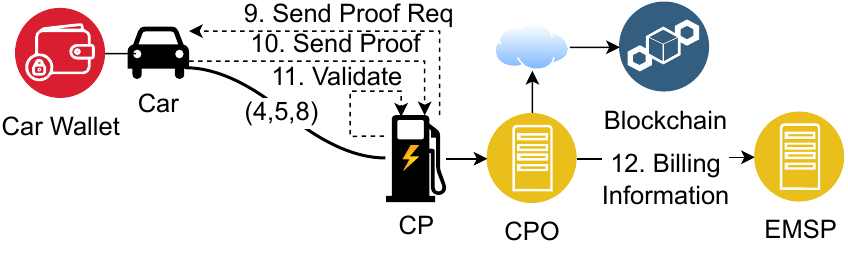}
 \caption{Credential Validation during the Charging Process}
 \label{fig:concept:credentialvalidation}
\end{figure}

The following section will outline the changes to the charging process (cf. \cref{fig:concept:credentialvalidation}). Specifically, the message sequence \textit{Identification, Authentication, and Authorization} from ISO~15118 is considered.

\paragraph{Step 9} \cref{fig:concept:credentialvalidation} shows the authentication of the vehicle by the \ac{CP}. 
In ISO~15118, service parameters such as the payment method are negotiated in the \textit{Service\-Discovery\-Req/-Res}. The authentication method now 
becomes another service parameter, making \textit{Contract Proof Identification Mode} a third option besides the existing modes (e.g., \ac{PnC}). In this \textit{Contract Proof} message sequence, \textit{Identification, Authentication, and Authorization} 
messages from ISO~15118 are changed after the \textit{PaymentServiceSelectionRes}.

By sending a \textit{RequestProofReq/-Res} the \ac{EV} receives a proof request from the \ac{CP}.
The \ac{CP}'s proof request includes a nonce and specifies which individual credential attributes the \ac{CP} expects in its role as verifier, not necessarily all the credentials/attributes issued to the \ac{EV} by the \ac{EMSP}.


\paragraph{Step 10} From the proof request, the \ac{EV} then creates a zero-knowledge proof for the requested attributes and a proof of non-revocation using its credential master secret and the \ac{CP}'s nonce. 
The proofs guarantee to a verifier that the \ac{EV} possesses valid non-revoked credentials for the identified attributes.
The \ac{EV} additionally uses its symmetric contract key to authenticate its contract towards the \ac{EMSP} by generating an HMAC over a hash of the \ac{CP}'s \textit{proof request}, a contract identifier, and a timestamp.
The hashed \textit{proof request} is used to bind the contract authentication data to the current \ac{CP}/session, the contract identifier is used by the \ac{EMSP} to identify the correct contract and symmetric contract key, and the timestamp is used to prevent replays.
The contract authentication data is encrypted for the \ac{EMSP} and sent together with the proofs to the \ac{CP} in a \textit{ValidateContractProofReq} message.

\paragraph{Step 11} The \ac{CP} can validate the zero-knowledge proof for the credential attributes by using the \ac{EMSP}'s public key and can validate the revocation status of the corresponding credential by using the \ac{EMSP}'s public tails file and the corresponding accumulator value from the ledger.
If all verifications are successful, the \ac{CP} responds with a \textit{ValidateContractProofRes} to the \ac{EV}.
Thereupon, the charging process can continue as described in ISO~15118.

\paragraph{Step 12} Finally, the encrypted contract authentication data (along with other billing relevant data, e.g., meter values) is sent from the \ac{CP} to its \ac{CPO}, who can forward this data to the corresponding \ac{EMSP}.
The \ac{EMSP} can decrypt the contract authentication data and identify the correct contract.
Hence, the usual billing relations are still possible, i.e., the \ac{CPO} can bill the \ac{EMSP} and the \ac{EMSP} can bill the \ac{EV} user.
However, the \ac{CP}/\ac{CPO} can no longer identify the specific \ac{EV} user and the \ac{EMSP} can no longer identify the specific charging location.


%


\section{Implementation}\label{sec:implementation}
To demonstrate the feasibility of the concept, the contract authentication described therein was implemented during the charging process together with all preceding initiation steps such as the creation of the \acp{DID} or the installation of contract credentials. 
Our implementation is based on the ISO~15118 reference implementation \textit{RISE-V2G} \cite{riseV2G}.
In order to compare the concept with the actual state of the standard, we compare our implemented methods with the default \textit{RISE-V2G} implementation.

The reference implementation covers all necessary features to establish comparability to the status quo and at the same time serve as a basis for the implementation of the concept. 
The project Hyperledger Indy\footnote{Hyperledger, 2021, \url{www.hyperledger.org}}
provides an implementation for all necessary \ac{SSI}-operations, thus the Indy SDK\footnote{Indy SDK, 2021, \url{github.com/hyperledger/indy-sdk\#libindy-wrappers}} was chosen to be integrated into our prototype. The reference implementation was extended by the steward and the \ac{EMSP} in addition to the existing services \ac{EV} and \ac{CP}. They are responsible for the detailed handling of the schemas, credential definitions, and credentials and interact with the other actors.
Our prototypical implementation focuses on the message sequence \textit{Identification, Authentication, and Authorization} and the associated communication between \ac{EV} and the other services as described in the concept. The actual accounting and communication between the secondary actors is not part of the implementation, as this is not in the scope of ISO~15118.
Additionally, the \ac{EMSP} onboarding, its creation of the three data structures \textit{Credential Schema}, \textit{Credential Definition}, and \textit{Revocation Registry} for the credentials of its customers' contracts and installation of \textit{Provisioning \ac{DID}} are also realized in the implementation.

The concept provides for the \ac{EMSP} to use the secure channel established by the exchanged \ac{DID} to create a \textit{WriteVerinymReq}. In the prototype implementation, however, communication is still secured via the old certificate infrastructure, as this has only been extended to include the \ac{EV} authentication. The \ac{CP} continues to authenticate itself via certificates.

\section{Evaluation}\label{sec:evaluation}

In this section, we evaluate the proposed/implemented solution.
Specifically, in \cref{sec:evaluation:p} we discuss the performance results based on our implementation from \cref{sec:implementation}, 
in \cref{sec:evaluation:t} we describe our Tamarin-based symbolic security and privacy proofs, 
and in \cref{sec:evaluation:r} we discuss how the concept addresses the defined requirements from \cref{sec:req}.

\subsection{Performance Measurements}\label{sec:evaluation:p}
Regarding performance, we evaluate the computational- and communication overhead of the proposed solution in comparison to the default ISO~15118 processes as implemented by \textit{RISE-V2G}.
For both types of overhead, the main changes are within the credential installation and charge authorization processes.
Details are shown in \cref{tab:messagestats:chargingsession}.

\begin{table*}
	\caption{Duration and Size of Charging Session Messages for both Implementations}
	\label{tab:messagestats:chargingsession}
	\centering
	\begin{tabular}{l p{1em} ll p{1em}  ll}
		\toprule
		\multicolumn{1}{l}{Message Name\hspace*{0em}}   & & \multicolumn{2}{c}{RISE-V2G}  & &  \multicolumn{2}{c}{SSI Impl.} \\
		& & time [ms] & size [byes] & & time [ms] &  size [byes]\\
		\midrule
		
		\multicolumn{5}{l}{\textbf{Credential Installation}}  & & \\
		\hspace{1em}CertificateInstallationReq         & &   296.0&811    & &     -&-      \\
		\hspace{1em}CertificateInstallationRes         & &   32.8&3638     & &     -&-      \\
		
		\hspace{1em}GetCredOfferReq                 	& & -&-  	& & 4.0&106           \\
		\hspace{1em}GetCredOfferRes                 	& & -&- 	& & 44.613&6710   \\
		\hspace{1em}CreateContractCredentialReq   		& & -&- 	& & 134.429&2185       \\ %
		\hspace{1em}CreateContractCredentialRes 		& & -&- 	& & 2603.864&5961   \\
		
		\multicolumn{5}{l}{\textbf{Charge Authorization}}  & & \\
		\hspace{1em}PaymentDetailsReq                  & &   649.8&1452    & &     -&-      \\
		\hspace{1em}PaymentDetailsRes                  & &   73.6&37     & &     -&-      \\
		\hspace{1em}AuthorizationReq                   & &   129.6&13    & &     -&-      \\
		\hspace{1em}AuthorizationRes                   & &   7.5&15      & &     -&-      \\
		\hspace{1em}RequestProofReq                    & &     -&-      & & 65.3&58   \\
		\hspace{1em}RequestProofRes                    & &     -&-      & & 3.6&266    \\
		\hspace{1em}ValidateContractProofReq           & &     -&-      & & 282.302&7281  \\ 
		\hspace{1em}ValidateContractProofRes           & &     -&-      & & 136.3&55  \\
		\bottomrule& & 
	\end{tabular}
\end{table*}

The communication overhead of the proposed solution for credential installation messages is 14,962 bytes in total. 
The default \textit{RISE-V2G} method requires 4,449 bytes for credential installation. 
Regarding charge authorization, the messages of the proposed solution are 7,660 bytes in total and the messages of the default \textit{RISE-V2G} method are 1,517 bytes. 
For comparison, based on our measurements, the total communication overhead of a full 1-hour default \textit{RISE-V2G} charge session with a credential installation and a charge status message interval of 10 seconds is roughly 20,000 bytes.
Hence, we argue, that the increased overhead of the proposed solution is still acceptable.

For computational overhead, all measurements were performed 1000 times\footnote{The measurements were performed on a Lenovo Thinkpad T480 with Intel® Core™ i5-8250U CPU @ 1.60GHz × 8, 15.5 GiB Ram, running Ubuntu 20.04.3 LTS 64-bit} and we report the respective average times (always including processing and message transfer). 
Regarding credential installation, the mean time of the proposed solution was 2786.9 ms compared to 328.8 ms with the default \textit{RISE-V2G} method. 
Regarding charge authorization, the mean time of the proposed solution was 487.502 ms compared to 860.5 ms with the default \textit{RISE-V2G} method (mostly due to certificate path validations).
The results show good performance for the proposed method, especially considering that credential installation is rarely performed (only if a new contract is concluded or old credentials renewed).

\subsection{Security and Privacy Analysis with Tamarin}\label{sec:evaluation:t}

We analyze the security of the proposed solution in the symbolic model using the Tamarin prover \cite{meier2013tamarin} and the corresponding files are provided online.\footnote{\label{f:git}\Git}
Tamarin is a state-of-the-art tool for automated security protocol analysis.
By default, analysis is performed in the symbolic model, i.e., assuming a Dolev-Yao adversary \cite{dolev1983security} with full control over the network who cannot break cryptographic primitives without knowing the respective private key (cf. adversary model in \cref{sec:req:attackermodel}).

With Tamarin, protocols are specified using a set of \emph{rules}, which define all relevant communication and processing steps of the protocol.
Additionally, security requirements are defined as trace properties (lemmas), which need to hold for all possible execution traces of the protocol, i.e., all traces that can be built with the defined rules.
Tamarin performs an exhaustive search for a trace that violates the defined requirements.
If a trace is found, this trace serves as a counterexample (a specific attack path that violates the requirement).
If no trace is found, the security requirement is proven to be satisfied by the defined protocol.

Furthermore, Tamarin enables the verification of observational equivalence properties, which can be used to show that an adversary cannot distinguish between two protocol runs. Observational equivalence is especially useful in order to verify privacy properties, e.g., by proving anonymity in \ac{EV} charging by showing that an adversary cannot distinguish between two charge authorizations of different \acp{EV}.

\subsubsection{Security Proofs}\label{sec:evaluation:t1} \hfill \\
The security requirements from \cref{sec:req:sec} require authentication between different actors over different data. 
The most commonly used notion to prove strong authentication properties is defined in \cite{lowe1997hierarchy}, namely \emph{injective agreement} (preventing spoofing, replay, etc.).
This property is defined as follows:

\begin{definition}[Injective Agreement \cite{lowe1997hierarchy}] \label{def:inj-agreement}
	A protocol guarantees to an initiator $A$ \emph{injective agreement} with a responder $B$ on a set of data items $ds$ if, whenever $A$ (acting as initiator) completes a run of the protocol, apparently with responder $B$, then $B$ has previously been running the protocol, apparently with $A$, and $B$ was acting as responder in his run, and the two agents agreed on the data values corresponding to all the variables in $ds$, and each such run of $A$ corresponds to a \emph{unique} run of $B$.
\end{definition}

Using our defined Tamarin model,\textsuperscript{\ref{f:git}} we successfully verify the following security properties based on the notion of injective agreement (cf. \cref{def:inj-agreement}).
For this, we assume one steward and the ledger is modeled as a secure storage, where only authorized entities can write but everyone can read. Communication with the ledger is assumed to be a secure channel as specifics of this communication are not part of our concept, but instead standardized by the respective ledger specification.
Additionally, we assume that the long-term key of all actors in a specific protocol run are secure since otherwise, attacks are trivially possible (e.g., if an \ac{EV}'s private provisioning key is leaked to an adversary, this adversary can spoof the affected \ac{EV} towards an \ac{EMSP} for contract credential installation).
However, in order to keep the needed
assumptions as weak as possible, other entities of the same types that are not
directly involved in the protocol run can be compromised.
For normal signatures/encryptions we use the built-in Tamarin functions.
The \ac{EV} zero-knowledge credential proofs are modeled with custom functions, whereby the \ac{EV} can create a zero-knowledge proof based on the installed credential and its master secret, which the \ac{CP} can verify with the \ac{EMSP}'s public key and revocation can be verified via a simple request over an accumulator in the ledger.
However, zero-knowledge proofs are modeled without specific cryptographic details, since, in the symbolic model, cryptographic functions are anyway assumed to be secure.
Besides the injective agreement-based lemmas to proof the desired security properties, our Tamarin files\textsuperscript{\ref{f:git}} also includes lemmas to verify the correctness of the defined model. That is, correctness lemmas are included to verify that the intended processes can be implemented with the defined rules and without adversary intervention in order to prevent the security properties from being trivially met by an incorrect model (e.g., all possible authentications are trivially secure if no authentication is possible at all).
In the following, we describe the verified security properties.
Note that the following paragraphs only provide intuitive descriptions of the verified properties as the full proofs are automatically generated with the Tamarin tool based on the defined models.
The full formal definitions are part of our Tamarin models (provided online\textsuperscript{\ref{f:git}} for reproducibility).

\definecolor{mygreen}{rgb}{0,0.5,0}
\definecolor{mymauve}{rgb}{0.58,0,0.82}

\lstdefinelanguage{tamarin}
{
	alsoletter={-}, 
	morekeywords={
		equations,
		functions,
		builtins,
		protocol,
		property,
		theory,
		begin,
		end,
		subsection,
		section,
		text,
		rule,
		pb,
		lts,
		exists-trace,
		all-traces,
		enable,
		assertions,
		modulo,
		default\_rules,
		anb-proto,
		in,
		let,
		Fresh,
		fresh,
		Public,
		public,
		restriction,
		lemma
	},
	keywordstyle=\color{mygreen},
	otherkeywords={
		=,
		@,
		<,
		>,
		|,
		],
		[,
		]-, 
		--, 
		==>,
		<=>,
		--[,
		]->,
		-->,
		.,
		\&,
		",
		:,
		\^,
		\$,
		!,
		~
	},
	morekeywords=[2]{
		hashing,
		h,
		asymmetric-encryption,
		aenc,
		adec,
		pk,
		signing,
		sign,
		verify,
		true,
		revealing-signing,
		revealSign,
		revealVerify,
		getMessage,
		symmetric-encryption,
		senc,
		sdec,
		diffie-hellman,
		inv,
		\^,
		1,
		bilinear-pairing,
		pmult,
		em,
		xor,
		zero,
		multiset
	},
	keywordstyle=[2]\color{teal}\textbf,
	morekeywords=[3]{
		=,
		@,
		<,
		>
	},
	keywordstyle=[3]\color{orange},
	morekeywords=[4]{
		Ex,
		All,
		F,
		T,
		|,
		.,
		\$,
		!,
		~,
		\&,
		==>, 
		<=>
	},
	keywordstyle=[4]\color{mymauve},
	morecomment=[l]{//}, 
	morecomment=[s]{/*}{*/}, 
	commentstyle=\color{teal},
	morestring=[b]', 
	stringstyle=\color{mymauve},
	breaklines=true,
	tabsize=2,
	xleftmargin=2em,
	basicstyle={\small}
}

\lstset{language=tamarin}
\begin{figure}[h]
\begin{lstlisting}[numbers=left, caption={Injective Agreement Lemma in Tamarin}, captionpos=b, label={fig:injAgreementEx}]
lemma auth_emsp_steward_verinym:
"All Steward S_DID EMSP Verinym_DID #i .
	CommitStewardVerinym(Steward, S_DID, EMSP, Verinym_DID) @i
	==> ( Ex #j . 
			RunningEMSPVerinym(EMSP, S_DID, Verinym_DID) @j
			& (#j<#i) 
			& not( Ex Steward2 EMSP2 S_DID2 #i2 .
				CommitStewardVerinym(Steward2, S_DID2, EMSP2, Verinym_DID) @ i2 
				& not(#i2=#i) ) )
		| ( Ex RevealEvent Entity #kr .
					KeyReveal(RevealEvent, Entity) @ kr 
					& Honest(Entity) @ i)"
\end{lstlisting}
\end{figure}

\paragraph{Secure Setup (\ac{EMSP} to steward)}
Regarding the secure setup (\ref{req:iss:trust}), we verify that an \ac{EMSP} and a steward (identified by their \ac{DID}) injectivly agree on the \ac{EMSP}'s verinym \ac{DID} (and corresponding public key) during the onboarding process. That is, whenever a steward $S$ accepts an \ac{DID} for writing on the ledger, apparently from an \ac{EMSP} $E$, $E$ has previously sent this \ac{DID} to $S$ and both actors agree on the content of the \ac{DID}. Additionally, each accepted \ac{DID} by $S$ corresponds to a unique request from $E$. The only allowed exception is, if the long-term key of one of the parties involved in a specific protocol run was leaked.

The Tamarin lemma, which models the \textit{Secure Setup (\ac{EMSP} to steward)} security property is shown as an example in \cref{fig:injAgreementEx}.
Hereby, lines 2--6 indicate that for every accepted \ac{EMSP} verinym \ac{DID} by as steward (identified by S_DID) at time \texttt{i}, there exists an event where the same \ac{EMSP} has sent this verinym \ac{DID} to the same steward at time \texttt{j} and \texttt{j} was before \texttt{i}.
Lines 7--9 models the uniqueness property of the acceptance by the steward, i.e., it says that there cannot exist another protocol run between the same or different actors (steward2 and EMSP2) where the same verinym \ac{DID} is accepted.
Lines 10--12 model the exception, that the security property can be broken if the long-term keys of one of the actors involved in the protocol (i.e., the actor was assumed honest at time \texttt{i}; line 12) run was revealed.

\paragraph{Secure Setup (cont.)}
Regarding the secure setup (\ref{req:iss:trust}), we additionally verify that a steward and an \ac{EMSP} injectivly agree on the steward's \ac{DID} public key during the onboarding process.
Furthermore, we verify mutual injective agreement between \ac{OEM} and steward during the onboarding process of an \ac{OEM} (see the full Tamarin models\textsuperscript{\ref{f:git}} for details).

\paragraph{Secure Contract Credential Installation}
Regarding the secure credential installation (\ref{req:issue:trust}), we verify that an \ac{EV} and an \ac{EMSP} (identified by their \ac{DID}) injectivly agree on a contract credential request and response respectively during the installation process (see the full Tamarin models\textsuperscript{\ref{f:git}} for details).
The only allowed exceptions are:
\begin{enumerate*}[label=\emph{(\roman*)}]
	\item if the long-term key of one of the parties involved in a specific installation protocol run was leaked or
	\item if the long-term keys of a previous \ac{OEM} to steward setup were leaked.
\end{enumerate*}

\paragraph{Secure Charge Authentication and Authorization}
Regarding the secure charge authentication (\ref{req:cont:trust}), we verify that an \ac{EV} and a \ac{CP} injectivly agree on an \ac{EV}'s charge request during the authentication process.
Additionally, for secure charge authorization/billing (\ref{req:cont:payment2}), we verify that an \ac{EV} and an \ac{EMSP} injectivly agree on an \ac{EV}'s charge authorization data for the billing process (see the full Tamarin models\textsuperscript{\ref{f:git}} for details).
The only allowed exceptions are:
\begin{enumerate*}[label=\emph{(\roman*)}]
	\item if the long-term key of one of the parties involved in a specific installation protocol run was leaked or
	\item if the long-term keys of a previous \ac{OEM} to steward setup were leaked or
	\item if the long-term keys of a previous credential installation were leaked.
\end{enumerate*}

\subsubsection{Privacy Proofs}\label{sec:evaluation:t2} \hfill \\
For our privacy analysis, we mainly focus on the verification of unlinkability properties based on \cite{brickell2009simplified} as previously used for the \ac{EV} charging context by \cite{kern2022integrating}.
Specifically, we use Tamarin to prove observational equivalence between two protocol runs that may be initiated by the same \ac{EV} or by different \acp{EV}.
Our models assume \textit{Honest-but-Curious Operators} (cf. adversary model in \cref{sec:req:attackermodel}) and we use separate Tamarin models per property for simplicity.
The following descriptions provide an intuitive description of the verified properties and full formal definitions can be found as part of the provided Tamarin models.\textsuperscript{\ref{f:git}} 

\paragraph{Non-Traceability}
Regarding preventing the creation of a movement profiles (\ref{req:char:traceability}), we verify unlinkability of \acp{EV}/users based on their billing relevant data (as received by the backend).
Specifically, we show that for two honest \acp{EV} $EV_1$ and $EV_2$, an adversary cannot distinguish between the scenario where charge billing data is received for an (authorized) session of $EV_1$ and $EV_2$ each and the scenario where charge billing data is received for two (authorized) session of $EV_1$.
Charge session may be at the same or different locations to show that linkability across locations (i.e., traceability) is not possible.

\paragraph{Non-Linkability}
Regarding the non-linkability of \ac{EV} users (\ref{req:char:linkability}), we verify unlinkability of \acp{EV}/users based on their authentication/authorization data (as generated by the \ac{EV}).
Analogously to non-traceability, we show that an adversary cannot distinguish between a scenario with two authorizations of different \acp{EV} and a scenario with two authorizations of the same \ac{EV}.

\subsection{Discussion of Requirements}\label{sec:evaluation:r}

The functional requirements are addressed by the concept design as follows:
Credential installation and charge authorization are still possible without user interaction \ref{requirement:interaction}, which ensures user-friendliness.
Contract authentication via \ac{SSI} can be negotiated via the \textit{ServiceDiscoveryReq/-Res} messages \ref{req:optional}.
All \ac{SSI} roles are covered by actors from the ISO-15118 ecosystem as discussed in \cref{sec:concept:conceptoverview} \ref{req:roles01}.
Vehicles manage their contract credential in their own wallet \ref{req:wallet}.
All contract credentials contain the same core elements as discussed in \cref{sec:concept}, which allows a \ac{CP} to authenticate the contract of different \acp{EMSP} \ref{req:iss:schema}.
Credential installation can be delayed until the first charging session \ref{req:installtime} using the messages described in \cref{sec:concept:contractissue:installationprocess}.
Communication of the \ac{EV} (e.g., for credential installation or reading data of the ledger) can still be tunneled via the \ac{CP} \ref{req:comm:tunnel} using the same concepts as for the default ISO~15118 method (e.g., credential installation messages are simply forwarded to the backend in Base64 encoding via OCPP 2.0 \cite{ocpp2-2}).
We judge the additional overhead to be acceptable \ref{req:perf} as discussed in \cref{sec:evaluation:p}.

The security requirements \ref{req:iss:trust}--\ref{req:cont:payment2} are addressed as discussed in \cref{sec:evaluation:t1}.
In short, the security requirements are shown to be met via symbolic proofs using the Tamarin tool.
The corresponding models for automated proof generation are provided online.\textsuperscript{\ref{f:git}}
All properties are verified in roughly 30 minutes on a standard laptop.\footnote{Using a Lenovo ThinkPad T14 Gen 1 with 16GB RAM.}
The published repository includes the defined model/lemmas, the used oracles (for performance such that the model analysis terminates within a reasonable time frame), and instructions on running the models (for reprehensibility of the formal analysis).

Analogously, the privacy requirements \ref{req:char:traceability} and \ref{req:char:linkability} are addressed as discussed in \cref{sec:evaluation:t2} and the models for automated proof generation are provided online.\textsuperscript{\ref{f:git}} 
The concept primarily prevents linkability/traceability through the authentication process at the \ac{CPO}/\ac{CP}. 
However, since traditional payment channels are still supported and thus charging sessions must be associated by the \ac{EMSP} with the respective customers, the \ac{EMSP} can still link them. This could be fixed via anonymous payment methods, which is out-of-scope for this paper.

\section{Conclusion}\label{sec:conclusion}

In this paper, we propose an approach for using \acp{SSI} as trusted credentials for \ac{EV} charging authentication and authorization in ISO~15118. 
By using verifiable credentials with zero-knowledge proofs, our solution addresses the privacy problems of ISO~15118 providing unlinkability of charging 
sessions. Furthermore, our solution uses a decentralized distributed ledger and does not require a complex centralized \ac{PKI} anymore.
Our prototypical implementation and performance evaluation show that the computational and communication overhead of our solution is relatively low
and should be acceptable for a real-world implementation. 
Our formal analysis using Tamarin shows that all required security and privacy properties hold, i.e., still guarantee authentication 
properties between different actors while preserving the \ac{EV} user's privacy to the highest possible extent (only \ac{EMSP} can link a user's 
charging events for billing purposes).
Future work could expand our concept to the authentication of all \ac{PnC} actors, especially \acp{CP}. 

\ifdefined\anonSubmission

\else
\subsubsection{Acknowledgements}
This research work has been partly funded by the German Federal Ministry of Education and Research and the Hessian Ministry of Higher Education, Research, Science and the Arts within their joint support of the National Research Center for Applied Cybersecurity ATHENE and the Deutsche Forschungsgemeinschaft (DFG, German Research Foundation) – project number 503329135.
\fi


\balance


\bibliographystyle{splncs04}
\bibliography{bibliography}

%
%

\end{document}